\documentclass[10pt,letterpaper]{article}
\usepackage[left=1.2in,right=1.2in]{geometry}

\usepackage{amsmath,amssymb}

\usepackage{changepage}

\usepackage{textcomp,marvosym}

\usepackage{cite}

\usepackage{nameref,hyperref}

\usepackage[nopatch=eqnum]{microtype}
\DisableLigatures[f]{encoding = *, family = * }

\usepackage[table]{xcolor}

\usepackage{array}

\newcolumntype{+}{!{\vrule width 2pt}}

\newlength\savedwidth

\newcommand\thickhline{\noalign{\global\savedwidth\arrayrulewidth\global\arrayrulewidth 2pt}%
\hline
\noalign{\global\arrayrulewidth\savedwidth}}

\usepackage[aboveskip=1pt,labelfont=bf,labelsep=period,justification=raggedright,singlelinecheck=off]{caption}

\makeatletter
\renewcommand{\@biblabel}[1]{\quad#1.}
\makeatother

\usepackage{graphicx}

\definecolor{ver}{rgb}{0.0, 0.5, 0.0}

\begin{document}
\vspace*{0.2in}

\begin{flushleft}
{\Large
\textbf\newline{Correction of Italian under-reporting in the first COVID-19 wave via age-specific deconvolution of hospital admissions} 
}
\newline
\\
Simone Milanesi\textsuperscript{1*},
Giuseppe De Nicolao\textsuperscript{2,3}
\\
\bigskip
\textbf{1} Department of Mathematics, University of Pavia, Pavia, Italy
\\
\textbf{2} Department of Electrical, Computer and Biomedical Engineering, University of Pavia, Pavia, Italy
\\
\textbf{3} Division of Infectious Diseases I, Fondazione IRCCS Policlinico San Matteo, Pavia, Italy
\\
\bigskip

%
%





* simone.milanesi01@universitadipavia.it

\end{flushleft}

\section*{Abstract}
When the COVID-19 pandemic first emerged in  early 2020, healthcare and bureaucratic systems worldwide were caught off guard and largely unprepared to deal with the scale and severity of the outbreak. In Italy, this led to a severe underreporting of infections during the first wave of the spread. The lack of accurate data is critical as it hampers the retrospective assessment of nonpharmacological interventions, the comparison with the following waves, and the estimation and validation of epidemiological models. In particular, during the first wave, reported cases of new infections were strikingly low if compared with their effects in terms of deaths, hospitalizations and intensive care admissions. In this paper, we observe that the hospital admissions during the second wave were very well explained by the convolution of the reported daily infections with an exponential kernel. By formulating the estimation of the actual infections during the first wave as an inverse problem, its solution by a regularization approach is proposed and validated. In this way, it was possible to computed corrected time series of daily infections for each age class. The new estimates are consistent with the serological survey published in June 2020 by the National Institute of Statistics (ISTAT) and can be used to speculate on the total number of infections occurring in Italy during 2020, which appears to be about double the number officially recorded.




\section*{Introduction}
The availability of accurate data on a pandemic outbreak is essential, as the data provide key information on the spread of the disease and enable authorities to assess and compare public health policies. Available data on COVID-19 cases can be used, e.g., for forecasting and analyzing deaths, hospitalizations, and occupation of intensive care beds, thereby deciding if and what containment measure to take. Also, estimating the size of an outbreak can predict the outbreak's future trajectory, thus informing planning about resources and interventions.

In Italy, the Istituto Superiore di Sanità (ISS) \cite{sito_dati}, was in charge of the official recording of COVID-19 data, based on which government committees were informed about the status of the pandemic phases and and advices were issued on non-pharmaceutical interventions (NPIs). It is widely recognized that, during the first outbreak, these official statistics largely underestimated the true number of cases  \cite{lau}. The Italian national Institute of Statistics (ISTAT) estimated the under-reporting during the first wave by conducting a seroprevalence survey published on \cite{istat}. It was found that the number of positive subjects was about six times larger than the number of registered cases. Knowing this ratio, however does not provide a reliable correction for the daily time series, apart from knowing that they suffer from a severe underdetection.

The issue of correcting for the underreporting has been addressed by several authors using a variety of approaches. Giordano and colleagues \cite{giordano} employed a compartmental model that incorporated the potential presence of undetected symptomatic and asymptomatic COVID-19 cases. The model's parameters were calibrated empirically using data on reported cases. Their simulations predicted that during the first phases as many as $35\%$ of the cases were not reported. Alternatively, also Bayesian methods have been employed \cite{wu}. Since they require a number of assumptions on the prior distributions of key parameters, these methods suffer from some subjectivity. 

An approach more grounded on evidence is to leverage the health effects of the outbreak in order to trace  back to the infections. For instance, whenever COVID-19 related deaths, hospitalization and ICU admissions or occupancy are less prone to underreporting, their recordings can be used to reconstruct the causes, i.e. the infections. Of course, this is possible provided that a reliable cause-effect model is available. For instance, in the delay-adjusted CFR (Case Fatality Ratio) model \cite{unnikrishnan}, deaths are modeled as a fraction of new cases delayed by some interval, see \cite{pullano}. In this case, the CFR  must be known in advance or estimated from a dataset where new cases do not suffer from underreporting \cite{thomas}. Correction of underreporting by the delay-adjusted CFR method is easily achieved letting the estimated time series of infections be equal to the time shift of the time series of deaths divided by the CFR. Of course, the same method can be used replacing deaths with other health effects such as hospital or ICU admissions or also hospital or ICU occupancy. In these cases, the cause-effect model consists of the fraction of severe cases leading either to hospital or ICU admission as well as the typical delay from infection to the admission. 

The main shortcoming of these delay adjusted methods is that they do not account for the nondeterministic nature of the delay between infection and the considered outcome, such as death or hospital/ICU admission. Indeed, the delay changes from patient to patient according to some probability distribution. 
In fact, due to the randomness of the delay, the time series of the outcome is not just a shifted version of the new cases, as recognized by Noh and Danuser, who devise an ingenious method, based on Expectation-Maximization, in order to improve the estimate of the ascertainment rate of new cases \cite{noh}. It is worth noting that the use of a unique Case Fatality Ratio, that makes no distinction for age, widens the confidence intervals of the reconstructed new cases.

A key observation, in order to achieve a better reconstruction of actual infections, is that hospitalizations could be better modeled as the convolution of new cases with a kernel function proportional to the random lag distribution. The proportionality coefficient is equal to the fraction of cases experiencing the given outcome: in the case of death it is the apparent CFR, and in the case of hospitalization it is the fraction of cases admitted. 

The implementation of more rigorous correction methods has been hampered by the need of enhanced approaches for both the identification and correction procedures. Specifically, in the identification phase, the estimation of the delay distribution must be taken into consideration in addition to the two parameters of delay and gain. This estimation could be performed utilizing epidemiological data not impacted by underreporting. Furthermore, in the correction step, scaling and shifting techniques need to be replaced by the resolution of an inverse problem, namely a deconvolution problem, which increases the complexity of the correction procedure.

In the present work, we address both steps: a technique based on linear system identification and regularization for inverse problems is proposed as a method to correct underreporting and used to reconstruct the time profile of new positives during the first COVID-19 wave in Italy. The availability of a rigorous reconstruction of the first phase of the pandemic is valuable under several respects. In particular, we notice that some retrospective analysis were conducted by using official and underestimated data, e.g., \cite{sebastiani}, which makes problematic the assessment of public health interventions, including lockdowns and vaccination campaigns. For instance, without knowing the fraction of susceptible subjects it is impossible to run counterfactual simulations relative to NPIs and vaccination deployment. Our correction method is believed to prove useful also in connection with studies of the early and cryptic transmission \cite{cauchemez} and those employing mobility data to study the spatial spread \cite{balcan}: in both cases the reliability of the results would be compromised if heavily underreported cases were input in the models.

The paper is organized as follows. In the \textit{Material and Methods} section, we introduce the system identification technique used to learn the cause-effect relationship between new positive cases and hospitalizations. We then describe the deconvolution problem and the algorithm for its solution. Finally, we present a method for assessing the uncertainty of the proposed reconstruction.
The \textit{Results} and \textit{Discussion} sections present our main findings and their interpretation, as well strengths and limitations of the proposed approach. The \textit{Conclusions} end the paper.

\section*{Materials and methods}
\subsection*{Data}
Data were downloaded from a publicly available platform (see \textit{Data Availability Statement}). A centered seven-day moving average was applied \cite{sartor}
to filter noise errors in the data, mostly due to weekly oscillations but also caused by random delays and recording errors.
\subsection*{Identification of input-output model of hospital admissions}
Letting the $t$ (days) be the integer time, $u_t$  will denote the observed time series of the daily number of  new positive subjects (at swab time) and $y_t$ the number of daily hospital admissions. It is assumed that $y_t$ obeys a convolution model:
$$
y_t = \sum_{k=0}^{\infty} g_ku_{t-k-D} + \epsilon_t,\quad
$$
where $D$ is an integer delay, $\epsilon_t$ is a zero-mean white noise and the kernel $g_t$ is the impulse response of the linear system with input $u_t$ and output $y_t$.

A useful constant related to the impulse response is the gain
$$
\mu := \sum_{t=0}^\infty g_t
$$
which represents the expected fraction of observed positive subjects that will eventually be hospitalized. The normalized function $\tilde g_t=g_t/\mu$ can be interpreted as the probability distribution of the random time $T_h$ elapsed from D days after the swab day to the day of hospital admission.
A particular case is given by an exponential distribution:

\begin{equation}\label{imprespexp}
g_t=\begin{cases} 0 & \mbox{if }t<0 \\
\beta\alpha^{t} & \mbox{if } t\ge 0
\end{cases}
\end{equation}
where $\beta>0$ and $0 \le \alpha <1$. The average of $T_h$ is $\bar T_h=\mathbb{E}[T_h]= \log(\alpha)$. Note also that 
$$
\mu =\frac{\beta}{1-\alpha}
$$

In order to estimate the parameters $\alpha$ and $\beta$ of the impulse response, the oe.m function of the MATLAB System Identification Toolbox was employed \cite{sysid}. 

\subsection*{Reconstruction of new cases via deconvolution}
An inverse problem is the process of estimating from a set of observations the unknown causal factors that produced them. In the context of the present paper, the set of observations are the hospital admissions, while the unknown causal factor is the time series of new cases. When the cause-effect relationship is described by a linear time-invariant system, the output signal is the convolution of the input with an impulse response and the inverse problem is called deconvolution. Observing that during 2020 new variants were not observed in Italy, it is assumed that the impulse response did not change from the first to the second wave.

In order to formulate the deconvolution problem, it is worth introducing a matrix representation of the convolution model. Let $U=(u_{t-D})_{t=1\dots n}\in \mathbb{R}^n$, $Y=(y_t)_{t=1\dots n} \in \mathbb{R}^n$ be the vectors of new cases and hospital admissions during the $n$ days of the first wave. Moreover, $E=(\epsilon_t)_{t=1\dots n}\in \mathbb{R}^n$ will denote the vector of the errors. 

Assuming $u_t=0, t\le -D$, if the impulse response $g_t$ is as in (\ref{imprespexp}), we have that
\begin{eqnarray}
    Y&=&GU + E \\
    \label{Gnovirtual}
    G&=&\left[\begin{array}{cccc}
    g_0 & 0 & \ldots & 0 \\
    g_1 & g_0 & & \vdots \\
    \vdots & & \ddots & 0 \\
    g_{n-1} & g_{n-2} & \ldots & g_0\end{array}\right]
\end{eqnarray}

Then, for a known impulse response $g_t$, the problem of estimating the new cases from the hospitalization admissions can be written as
\begin{eqnarray}
\label{prob_min}
	C= \arg\min_U \{ \phi(Y,GU) +\lambda J(U) \}
\end{eqnarray}
where 
\begin{description}
	\item $C \in \mathbb{R}^n $ is the vector of estimated new cases;
	\item $G\in\mathbb{R}^{n\times n}$ is the convolution matrix obtained by the impulse response $g_t$ identified from the second wave data;
	\item $\phi: \mathbb{R}^n\times\mathbb{R}^n\to\mathbb{R}_{\ge0}$ is the loss function, whose purpose is to measure the data fitting;
	\item $J:\mathbb{R}^n\to\mathbb{R}_{\ge0}$ is a regularization penalty function, whose purpose is to measure irregularity of $U$; 
	\item $\lambda\ge0$ is the hyperparameter that adjusts the relative importance between of $\phi$ and $J$. 
\end{description}
In order to relax the assumption $u_t =0 , t\le -D$, we can include as additional unknowns the $L+1$ input values $u_t, -(D+L) \leq t \leq -D$, so that $U=(u_{t-D})_{t=-L,\ldots n}\in \mathbb{R}^{n+L+1}$
In this case, the matrix $G$ in (\ref{Gnovirtual}) has to be updated as follows
\begin{eqnarray*}
    G&=&\left[\begin{array}{cccccc} 
    g_{L+1} & \ldots & g_0 & 0 & 0 & 0 \\ g_{L+2} &\ldots &\ldots & g_0 & 0 & 0 \\ \vdots & & & & \ddots & 0 \\     g_{L+n} & \ldots & \ldots & \ldots & \ldots & g_0\end{array}\right]
\end{eqnarray*}
Although the number of unknowns is greater than the number of observations,  by a proper choice of the loss $\phi$ and the regularization term $J$ one can still ensure that (\ref{prob_min}) is convex so that a unique solution exists.

Concerning the cost function, a classical choice is to use a quadratic loss $\phi(v) = ||v||_2^2=\sum_{i=1}^n v_i^2$ and a regularization penalty  $J$ equal to the squared $2$-norm of a linear operator applied to the input, i.e. $J= ||Pu||_2^2$. Some possible choices for P are:
\begin{itemize}
    \item $ P=I_n \in \mathbb{R}^{n\times n}$ or ridge penalty;
    \item $P=\Delta_n \in \mathbb{R}^{n\times n}$ or penalty on first differences;
    \item $P=\Delta_n^2 \in \mathbb{R}^{n\times n}$  or penalty on second differences.
\end{itemize}
where $I_n$ is the identity matrix of order $n$ and
$$
\Delta_n =  
\begin{bmatrix} 
1 & 0 & \cdots & \cdots & \cdots & 0 \\ 
-1 & 1 & 0 & \cdots & \cdots & 0 \\
0 & -1 & 1 & 0 & \cdots & 0 \\
\vdots & &\ddots & \ddots & & \vdots \\
\vdots & & &\ddots & \ddots & 0 \\
0 & \cdots & \cdots & 0 & -1 & 1 

\end{bmatrix}
$$
Under these assumptions, (\ref{prob_min}) is a quadratic programming problem whose closed form solution is
$$ C= (G^TG+\lambda P^TP)^{-1}G^TY $$
Observe that there is no guarantee that $c_t\ge0 \textbf{ } \forall t=1 \dots n$. Whenever violated, the nonnegativity constraint is dealt with by reformulating the problem as a constrained quadratic programming one, solved via Matlab's Optimization Toolbox \cite{optim}. 

When the number of data and the number of unknowns are large, it may be convenient to consider a linear programming approach, e.g by letting
$$\phi(\cdot) = ||\cdot||_1, J(\cdot) = ||P(\cdot)||_1$$
where $||\cdot||_1,$ denotes the 1-norm, i.e.  $||v||_1=\sum_{i=1}^n |v_i|$. The linear programming problem is then equivalent to minimizing
$$\sum_{t=1}^n b_t+ \sum_{t=1}^n d_t$$ 
subject to the constraints
\begin{itemize}
    \item [(L1)] $b_t\ge y_t-(Gc)_t$
    \item[(L2)] $b_t \ge -y_t+(Gc)_t$
    \item[(L3)]$ d_t\ge\lambda(Pc)_t$
    \item[(L4)] $d_t\ge -\lambda(Pc)_t$
    \item[(L5)] $c_t,b_t,d_t \ge 0 \; \forall t=1\dots n $
\end{itemize}
where the decision variables represent $ c_i= C_i $, $b_t=| y_t-(Gc)_t|$ and $d_t=| \lambda (Pc)_t |$.

Finally, regardless of the selected approach, the tuning of the regularization penalty has to be properly addressed. Herein, Mallow's  $C_p$ criterion is adopted, assuming $\mathbb{E}[E]=0, \mathrm{Var}[E]=\sigma^2 I_n$ \cite{mallows}. The variance $\sigma^2$ is estimated by fitting an overparametrized model, i.e. a $20$-degree polynomial, to the hospital admission data.

\subsection*{Assessing uncertainty via boostrap }
In order to assess the uncertainty of the estimated impulse response, we resorted to a \emph{wild bootstrap} resampling scheme \cite{wucfj}, which  allows for heteroschedastic data, like the hospital admissions whose variance is possibly nonuniform. 

Concerning the uncertainty of the impulse response identified from the second wave data, the idea is to start from the predicted time series $\hat y_t$ of hospital admissions obtained by convoluting the new cases with the identified impulse response $\hat g_t$:
$$
\hat y_t = \sum_{k=0}^{\infty} \hat g_k u_{t-k-D}
$$
Then, letting $e_t=|y_t-\hat y_t|$, the bootstrapped datasets are obtained by generating resampled data as 
$$
y^*_t = \hat y_t + e_t v_t
$$
where $\{v_t\}$ are indipendent and identically distributed standard normal variables. Each boostrapped series of hospital admissions is then fed into oe.m to obtain a new realization $\hat g^*_t$ of the identified impulse response. Given a sufficient number of realizations, for any given time $t$, it is then possible to compute percentiles for $g_t$. Moreover percentiles for the gain $\mu$ and the time constant $\bar T_h$ can also be obtained. 

For what concerns the uncertainty of the reconstructed profile of first wave new cases, knowing the uncertainty of the impulse response is not sufficient, because a
second source of uncertainty must be accounted for, namely the noise affecting the daily hospital admissions during the first wave.
To account for both sources, a double wild boostrap scheme can be used: not only is the deconvolution replicated on the boosted hospital admissions but, in each replication, a different impulse response from the previously described wild boostrap performed on the second wave is used. As a  result, several realizations of the reconstructed new cases are obtained, so that percentiles can be computed both pointwise or relative to the total cases.



\section*{Results}
\subsection*{Hospitalization model during the second wave}
\begin{figure}[!ht]

\includegraphics[width=1\textwidth]{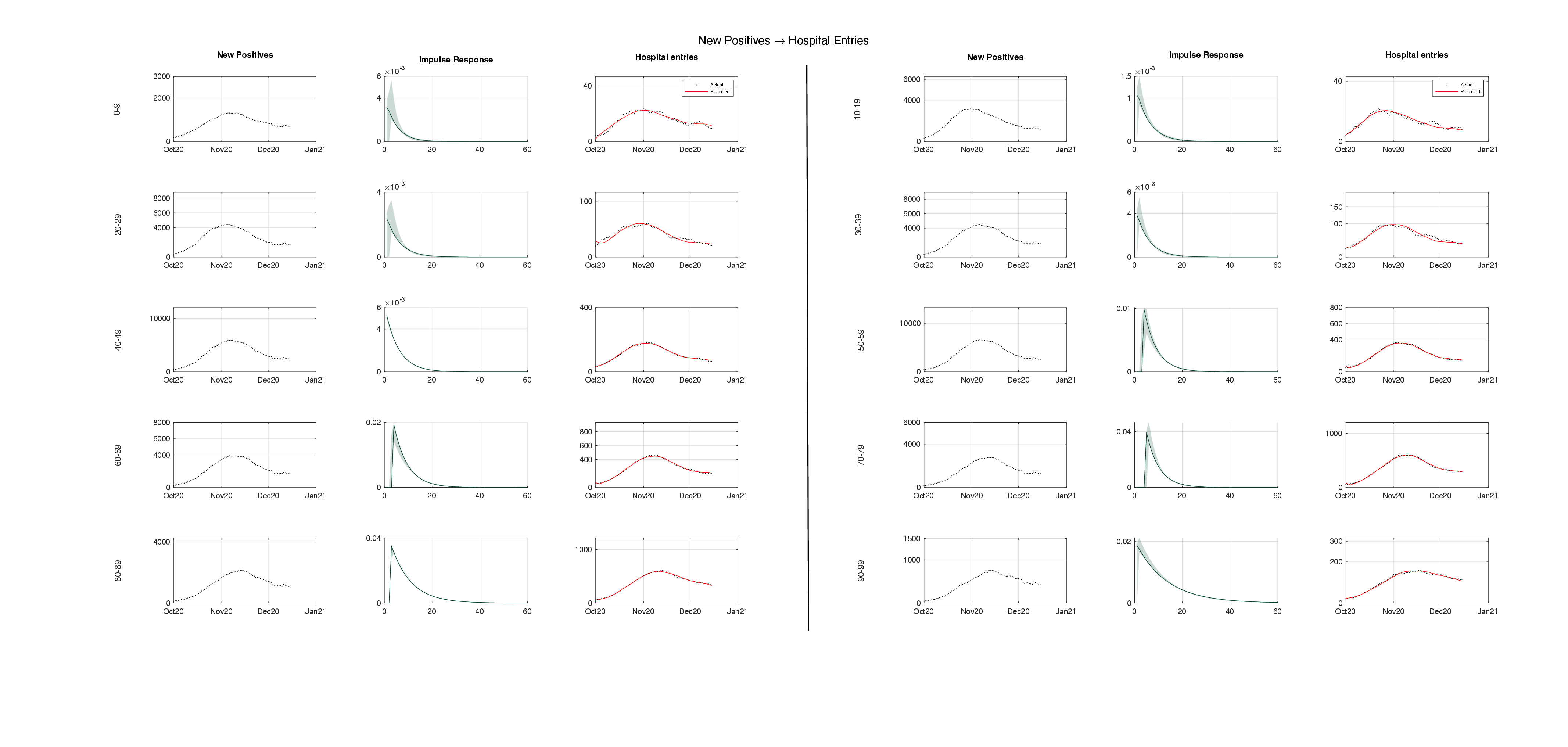}

\caption{{\bf Identification of input-output model of hospital admissions.}
In the first and fourth columns, the daily new cases (black dots) during the second wave  are plotted for the ten age groups. The new cases for all age groups showed an upward trend until early November, when the growth rate slowed down and a declining phase began in correspondence of the government's NPIs. An exponential impulse response model was identified for all age groups using Matlab's \textit{oe.m} function. The impulse responses (green) are shown in the second and fifth columns, along with pointwise confidence intervals calculated via wild bootstrap. The new cases were convoluted with the estimated impulse response to produce the  predicted hospital admissions (red) for all ten age groups that are plotted in the third and sixth columns together with the actual admissions (black dots).}
\label{fig1}

\end{figure}

The daily new cases and daily hospital admissions for the different age groups during the second wave in the period 1 October - 15 December 2020  are displayed in Fig \ref{fig1}. For all ages, the time series of the new cases exhibit an increase until the beginning of November when the progression slows down and a declining phase starts in correspondence with the containment measures taken by the Italian Government. The time series of the daily hospital admissions show a similar profile. Depending on the age group, their maxima range from some tens (e.g. age 0-9) to some hundreds (e.g. age 70-79). 

The exponential impulse response models and the delay parameters $D$ were identified for all age groups from the new positives and hospital admissions via Matlab's oe.m function. It resulted that the delay parameters $D$ were negative for all age groups and ranged from -10 to -6. This does not violate causality, because $D$ would be strictly positive is new cases were recorded at infection time, while the positive cases were actually recorded at swab time. The obtained impulse responses are displayed in Fig \ref{fig1} together with pointwise confidence bands computed via wild boostrap. The differences between the ten impulse responses can be appreciated from the upper panel of Fig \hyperref[S1Fig]{S1}, where they are displayed together.
Each exponential impulse response is characterized by two parameters: its area, i.e. the gain $\mu$, and the time constant $\bar T_h$, whose estimated values are reported in Table \ref{table1} together with their $2.5$th and $97.5$th percentiles. The gains ranged from $0.68\%$ for the age group $10-19$ to $30.62\%$ for the age group $80-89$, reflecting how illness severity varied with age. Concerning the time constant, it ranged from $4.8$ days (age $0-9$) to 12.8 days (age $90-99$). Both parameters were estimated with good precision. If impulse is normalized to have unit area, it can be interpreted as a probability density function of elapsed time. The ten density functions are displayed together in lower panel of Fig \hyperref[S1Fig]{S1} in the Supplementary Materials. The age groups until 49 year-old exhibit a very similar distribution with shorter mean compared to the three age groups from 50 to 79 year-old. Finally the last two age groups have the longest time constants.

In order to assess the convolution model, the new cases were convoluted with the estimated impulse response, thus  the time series of predicted hospital admissions for all the ten age groups. As seen in Fig \ref{fig1}, in all cases there is very good agreement between the predicted and observed items.

 \subsection*{Reconstruction of first-wave infections via deconvolution}
\begin{figure}[!ht]

\includegraphics[width=1\textwidth]{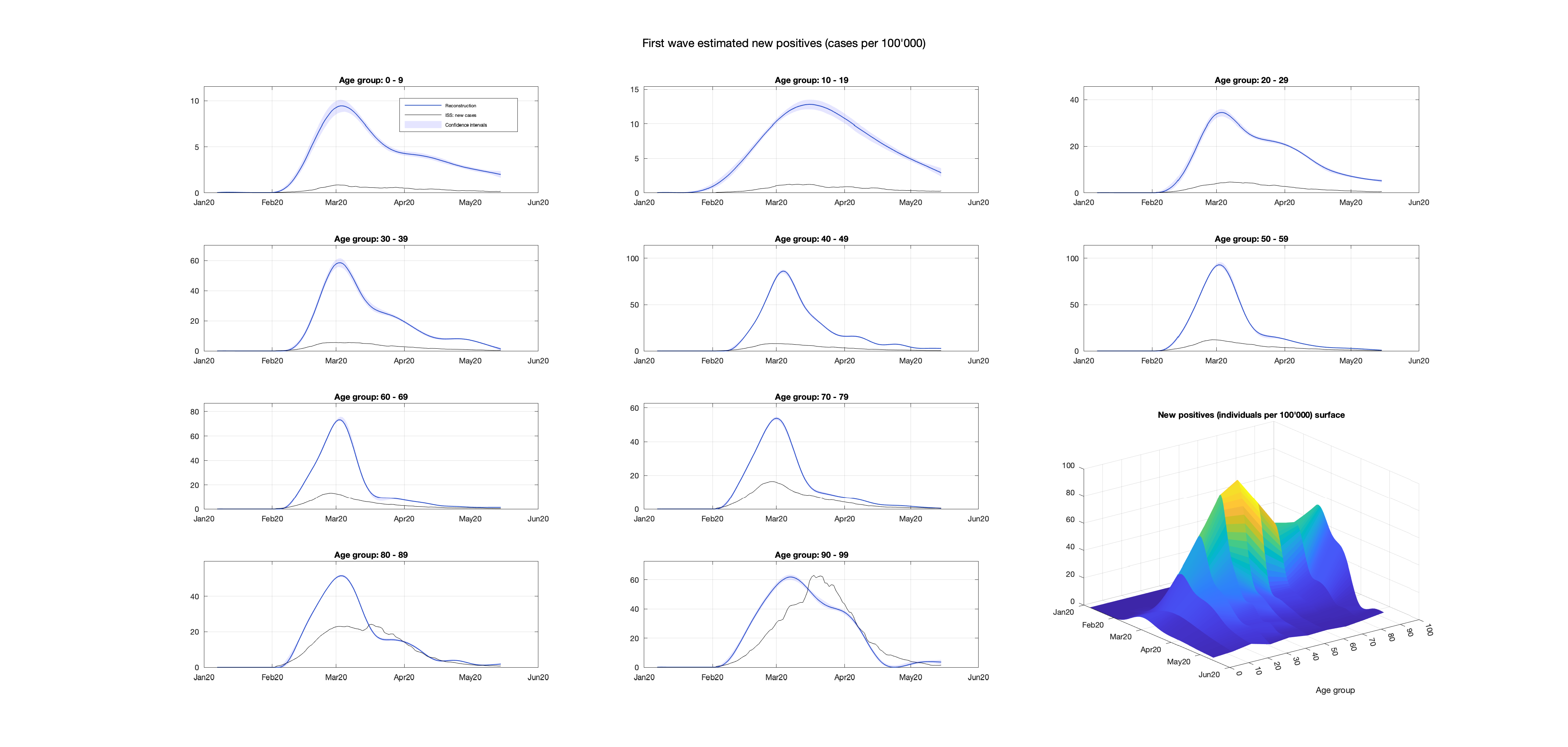}

\caption{{\bf Reconstruction of first-wave infections through regularized deconvolution.}
The new cases during the first wave were reconstructed via regularized deconvolution, starting from the daily hospital admissions. The estimated new cases per $100,000$ individuals (blues) are shown along with pointwise $95\%$ confidence intervals (light blue). The recorded cases (black) are also plotted for comparison. The patterns are similar across the age groups, featuring a rapid increase, a peak in early March, and a subsequent decline. In the bottom right corner, the ten profiles are combined to form a surface, providing the new cases per $100,000$ as a function of date and age.}
\label{fig2}
\end{figure}

The new cases during the first wave (7 January -  15 May 2020) were reconstructed via regularized deconvolution applied to the time series of first wave hospital admissions. The estimated new cases per $100,000$ subjects are displayed in Fig \ref{fig2}, together with pointwise $95\%$ bands. The shapes are similar with steep rise, a peak at the beginning of March and a subsequent decay. In all age groups the time series of the reconstructed new positives is uniformly larger than that of official ones, except for the age group 80-89, where official cases exceed the reconstructed ones in the second half of March, and the age group 90-99 where the phenomenon, occurring since mid March, is even more apparent. 
In the lower right corner of Fig \ref{fig2}, the ten profiles are plotted together to form a surface, giving new cases per $100,000$ as a function of date and age.

\begin{figure}[!ht]

\includegraphics[width=1\textwidth]{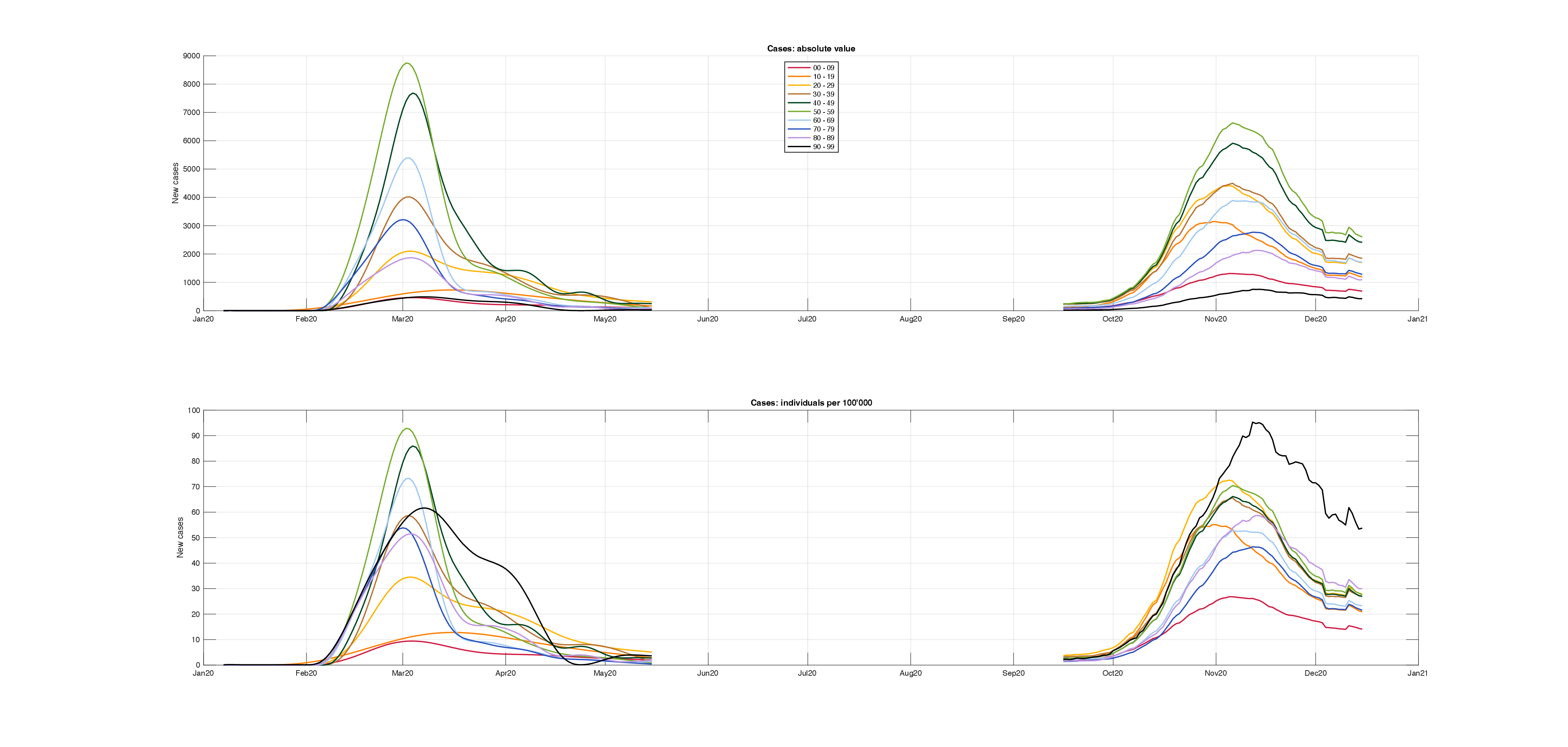}

\caption{{\bf Comparison between the reconstructed first-wave cases and official second-wave ones.}
A simultaneous view of all age groups is provided, where the upper panel shows the absolute values and the lower panel displays the cases per $100,000$ individuals. With regards to absolute numbers, the highest daily case count for both waves was among the age groups 40-59, with higher values seen in the first wave. When looking at cases per $100,000$, some differences between the two waves should be noticed. During the first wave, the 40-49 and 50-59 age groups had the highest peaks, while in the second wave the 90-99 and 20-29 age groups had the highest peaks. The 90-99 age group showed the slowest decline in both waves. The 0-9 and 10-19 age groups had the lowest peaks in the first wave, but in the second wave the 0-9 group had a higher peak of 27 daily cases per $100,000$. The peak value for the 10-19 group was early and exceeded 50 daily new cases, surpassing the 60-69 and 70-79 age groups.}
\label{fig3}
\end{figure}

Fig \ref{fig3} provides a simultaneous look at all the age groups for both absolute number (upper panel) and cases per $100,000$ (lower panel). Concerning absolute numbers, the largest numbers of daily cases referred, for both waves, to the age groups from 40 to 59, with higher values reached in the first wave. Concerning the cases per $100,000$, the two waves show some differences. In the first one, the first two peak values are again reached by the age groups  40-49 and 50-59, while in the second wave the first two peak values are reached by the 90-99 and 20-29 age groups. In both waves, the 90-99 group is the one exhibiting a slower decay. Another major difference regards the younger age groups. In the first wave the 0-9 and 10-19 groups show the lowest peak values,
below 10 and 15 cases per $100,000$, respectively. In the second wave, instead, the group 0-9 has still the lowest peak value, but reaches 27 daily cases per $100,000$, while the new cases of group 10-19, not only rise earlier but their peak value exceeds $50$
daily new cases, more than the peak values of the 60-69 and 70-79 age groups.

\begin{figure}[!ht]

\includegraphics[width=1\textwidth]{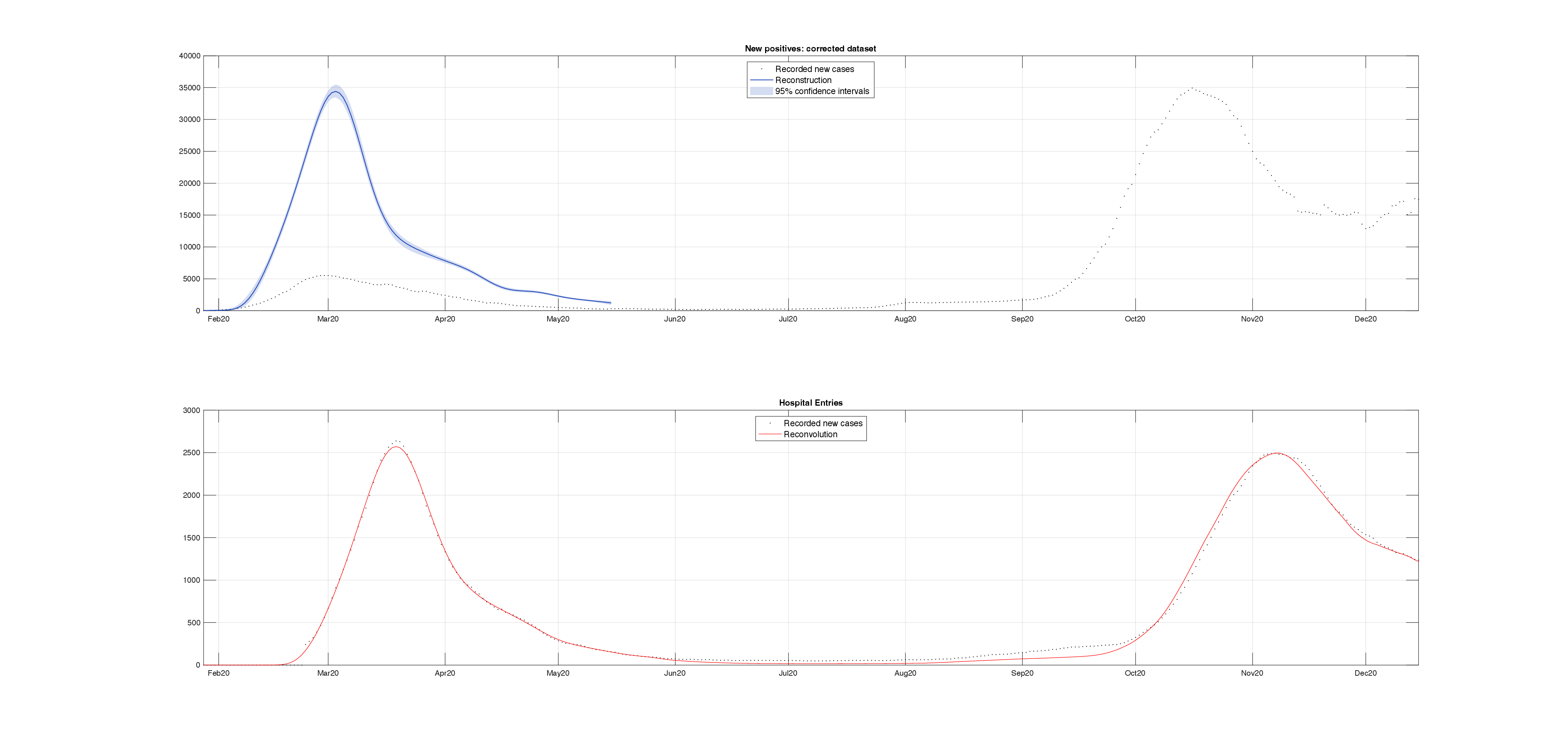}

\caption{{\bf Corrected and reconvoluted dataset.}
The estimated new cases for all age groups can be summed to get a corrected version of the overall new cases during the first wave. The upper panel compares the corrected series (blue) with the officially recorded new cases (black dots) from 7 January to 15 December, 2020. The analysis confirms that the official data significantly underestimated the number of new cases during the first wave, whose peak value was much lower than those observed during the second wave. The rise and decay of new cases during the first wave were steeper compared to the second wave, even though the peak values were similar. The lower panel shows that the hospital admissions (black dots) match well with the admissions predicted by the convolution model (red).}
\label{fig4}
\end{figure}

The estimated new cases for all age groups can be summed to obtain a corrected version of the global time series of the new cases during the first wave, displayed in Fig \ref{fig4}, together with the official recorded new cases from 7 January to 15 December. Thanks to the reconstruction procedure, is appears that the official data severerely underestimated the number of new cases during the first wave, whose peak values appeared much smaller than that observed during the second one. As a matter of fact, the reconstructed peak value reached at the beginning of March is about equal to the $35,000$ cases observed in mid October 2020. In spite of the similar peak values, the two waves have different slopes: both the rise and the decay of the new cases during the first wave are steeper. In the lower panel of Fig \ref{fig4}, it is shown that the hospital admissions (black dots) are well fitted by the admissions predicted by the convolution (red curve). The prediction of hospital admission for all age groups are displayed in Fig \hyperref[S2Fig]{S2}, where the fit appears good except for youngest age groups (up to 39 year-old) during the late summer period, while for older age groups the fit appears satisfactory even during that period.

It is of some interest to assess the daily values of the underestimation factor, defined as the ratio of the reconstructed number of new cases to the number of official ones.
It is seen that the daily underestimation factor was much larger than one, reaching a six-fold underestimation at the beginning of March, followed by a decrease until three and some further wandering around four. 

\section*{Discussion}
In this paper we addressed the underreporting of Italian COVID-19 cases during the first wave, by exploiting a convolution model of hospital admissions. A first assumption underpinning our study is that the second wave was not subject to a massive underreporting as that occurred during the first one. A second assumption is that in both the first and second wave, for each age-class, the hospital admissions were explained with good approximation by a convolution model involving the time series of the new cases and a suitable kernel function. Under these assumptions, the new cases during the first wave were reconstructed by solving two inverse problem. The first one, formulated on the second wave data, yielded the kernel function of the hospital admission model as the solution of a parameter estimation problem involving  hospital admissions and new cases. In the second step, the new cases during the first wave were obtained via deconvolution, using the first-wave hospitalizations and the kernel function estimated in the first step.

Concerning hospital admissions, the results showed that a simple 3-parameter kernel, consisting of a delayed and scaled exponential function, predicted remarkably well the hospital admissions of all age groups, thus validating the convolution model. The kernel parameter estimates shown in Table \ref{table1} seem meaningful: for instance, the kernel areas, representing the fraction of hospitalizations among the positive subjects, tend to increase with the age, ranging from a minimum of $0.7\%$ (10-19 year-old) to a maximum of $30.7\%$ (80-89 year-old). Also the delay between swab time and hospital admission as well as the time constant tend to be rather homogeneous across age groups. The negative delays, ranging from -6 to -10 days, are likely explained by different delays in the two pipelines of infection and hospitalization registries.

The solution of the second inverse problem yielded the reconstructed time series of daily new cases during the first wave. Differently from the first inverse problem, a simple parametric model describing the unknown signal was not available, so that the solution hinged on a machine learning regularization approach. The use of a penalty on the squared second differences proved effective. Alternative approaches, i.e. $\ell_1$-norm and ridge- and first-difference quadratic penalties, were also tried, but without improvement (data not shown).

As seen in Fig. \ref{fig2}, it turned out that, for all ages, the reconstructed profiles of daily new cases were definitely higher than the official data, thus confirming  the massive underreporting, highlighted by several authors REFS. For all age groups the peak value was reached almost simultaneously in the first half of March 2020. Apart from younger (until 29 year-old) and older (from 80 year-old) subjects, the steep rise of cases was followed by a quick decline during the enforcement of NPIs. Younger and older groups showed a somehow slower decline. In any case, the shape of the curves changed gradually between adjacent age groups, as seen in the surface plot in the lower right corner of Fig. \ref{fig2}.

The lower panel of Fig. \ref{fig3}, displaying new cases per 100,000, makes it possible to compare the impact of the first and second wave on the different age groups. A first observation is that during the first wave the time series of new cases exhibited high and narrow profiles, with an acute angle while the profiles during the second wave appeared definitely smoother. This change of behavior is consistent with the different severity of the containment measures that passed from strict national lockdown established by the Government on March 9, 2020, to a region-based and adjustable set of measures adopted during the second wave. Major differences between the waves were observed for the younger age groups until 29 year-old: during the most severe phase of the first wave they exhibited the lowest values, while in the second wave a different behavior was observed. In fact, the 0-9 age group, while still having the lowest peak, tripled its value. In the second wave, the 10-19 and 20-29 age groups, not only more than doubled their peak values but rised more quickly. These observations are consistent with early closure of schools and universities in the first wave \cite{wiki}, while closures were not so prompt during the second one. Another major difference between the two waves regards the oldest age group, i.e. 90-99. During the second wave the peak value was the highest one among all age groups and more than 1.5 times higher than the reconstructed peak value for the same age group during the first wave. However, our reconstruction is based on hospitalization data, so the profile of the first wave could be distorted if hospital bed eligibility changed from wave to wave for specific age groups, such as older individuals.\cite{cesari}.

A crucial issue regards the assessment of uncertainty affecting our estimates. In order to account for the eteroskedasticity of observed hospital admissions, uncertainty propagation was assessed by means of a wild boostrap scheme. It was found that the 3-parameter kernel was estimated with a very good confidence, see Fig. \ref{fig1} for a visual appraisal, and Table \ref{table1} for the $95\%$ confidence intervals of the kernel parameters, which were narrow for all age groups. As expected, wider ranges were observed for age groups with fewer hospitalizations.

The main idea behind our reconstruction scheme is modeling the effect of contagions on hospitalization during the second wave and then inverting this causal model to obtain the contagions during the first wave from the observed effect. As observed effect we used the hospital admissions, but other choices could be the ICU admissions or the deaths. There are several reasons suggesting that hospital admissions are to be preferred. First of all, the number of deaths and ICU admissions are definitely smaller, which, in view of the Poissonian nature of these occurrences yields noisier signals. Also, deaths may not be a completely reliable metric to compare the impact of the first and second waves of COVID-19 because the assumption of comparability may not be valid. In fact, not all deaths caused by COVID-19 during the first wave were accurately recorded as such \cite{modi}. Moreover, the average time elapsed from diagnosis to death is longer than that to hospitalization, so that the kernel function has a slower decay. As a consequence, the deconvolution problem becomes more ill-conditioned \cite{ekstrom}. In an analogous way, using hospital occupancy as observed effect, in place of admissions,  would entail a slower kernel function, worsening again the ill conditioning. Finally, concerning ICU admissions, it is also possible that during the first wave the availability of ICU beds reached saturation \cite{cesari, sartor, grasselli}, which would again violate the comparability assumption. Of course, some degree of saturation during the first wave may have occurred also for hospital beds, which implies that our estimates are somehow conservative, in the sense that the reconstructed cases, though much greater than official ones, might still underestimate the actual ones.

Our analysis led to three main findings. The first one regards the possibility of accurately predicting hospital admissions by means of a simple convolution model whose exponential kernel depends on three meaningful parameters, the scale factor, the delay, and the decay rate. From the scale factor and the decay rate, the percentage of positive subjects, for each age group, that are hospitalized can be derived as a secondary parameter. The prediction of the impact of COVID-19 on the healthcare system has emerged as a key issue since the early phase of the pandemic, see e.g. \cite{remuzzi}, where the fitting of ICU occupancy by means of an exponential curve was used to obtain a short term prediction of the strain on critical care facilities. Several authors derived simplified causal models that predict healthcare outcomes as delayed and scaled versions of the time series of the new cases \cite{unnikrishnan, pullano}. A more realistic modeling approach describes the outcomes as the convolution of the new cases with suitable kernel functions. This approach was pursued in \cite{giordano2} in order to assess the healthcare system costs associated with different vaccination rollout scenarios. Differently from the present paper, were hospital admissions have been modeled, in \cite{giordano2}, besides deaths, hospital and ICU occupancy were predicted. A possible future development may regard the identification of analogous models for the ICU admissions and the COVID-19 related deaths, as well as for hospital and ICU bed occupancy. Due to the convolution, these time series follow with varying delays the time profile of the new cases. Therefore the availability of reliable predictive models may help to take decisions and manage the healthcare resources, see e.g. the role played by disease severity parameters in COVID-19 Pandemic Planning Scenarios \cite{cdc}.

It should be noted that these predictive models are distinct from and complementary to classical epidemiological models, which are geared toward describing the evolution of the number of susceptible, infected, and recovered people in a population where the virus is spreading. Given the structural uncertainty related to the difficulty of predicting policy decisions and their effects on the spreading of the virus, it is inevitable to propose a range of scenarios with varying degrees of severity. It is precisely this variety of scenarios, which included worst-case catastrophic outcomes, that is at the origin of some controversy in the media about the models developed by Imperial College in England \cite{imperial} and the Kessler Foundation in Italy \cite{kessler}. In contrast, as shown in this paper, the day-to-day cause-and-effect relationship between numbers of infections and hospital admissions can be predicted with a high degree of reliability.

The second main finding regards the assessment of the underreporting ratio during the first wave, defined as the ratio between the actual infected subjects to the official cases. Our methodology makes it possible to equalize underreporting between the first and second wave. In particular, in order to restore comparability, the official cases of the first wave should be multiplied by a factor $4.5$. Moreover, as seen in Fig \ref{fig2}, such ratio factor did not remain constant during the first wave. This means that, without some form of correction, the data from February to May 2020 are unusable for almost all epidemiological purposes. It is worth observing that, even after equalization, there remains some underdetection, given that a significant fraction of cases were not recorder also during the second wave. Differently from UK, where the ONS monitored a statistical sample throughout the pandemic, the Italian national institute of statistics, ISTAT, carried out only one survey, just after the first wave \cite{istat}. According to this survey, the actual cases were $6.2$ times larger than the official ones. This means that, even after our reconstruction, a further $1.4$ fold underdetection  should be accounted for before arriving at the true number of cases. This suggests that also during the second wave at least one third of the cases were not recorded and the total number of cases during 2020 was about $4$ million. In other words it is estimated that about $7\%$ of the Italian population had been infected by the end of 2020 against the official $3.5\%$. This means that by the end of 2020, there were twice as many actual cases as officially registered ones.  For the sake of comparison, in early December 2020, the ONS estimated that in England about $7$ million people had antibodies, corresponding to around $13\% (10\%-16\%)$ of the English population \cite{spiegel}.

The third finding has to do with the comparison of age groups in the two waves. Indeed, it was found that some age groups exhibited notable differences. In particular, the 0-9 and 10-19 age groups had the lowest peaks in the first wave, below
$10$ and $15$ cases per 100, 000, respectively. In the second wave, instead, the group 0-9 had still the lowest peak value, but reached 27 daily cases per $100,000$, while the new cases of group 10-19, not only rose early but their peak value exceeded $50$ daily new cases, more than the peak values of the 60-69 and 70-79 age groups. These differences may be worth some further investigation, being associated with school reopening in September 2020, compared with early closure in Northern regions during the first wave \cite{wiki}. Another notable difference between the two waves concerns the time series of positive cases in the age group $90-99$: the peak value of daily cases per $100,000$ passed from 62 in the first wave to 95 in the second one. Before drawing any conclusions on the management of resthomes, it must be kept into account that our estimate may biased  by a change in eligibility for hospital beds for infected elderly individuals, motivated by saturation that occurred during the first wave. The issue could be explored further by replicating our deconvolution approach using deaths in place of hospital admissions, under the assumption that deaths related to COVID-19 were recorded with uniform criteria during the two waves.

\subsection*{Strenghts and limitations}
This study introduces a method to correct the underreporting of new cases during the first COVID-19 wave, taking into account the differences between age groups, so that the temporal evolution of the underreporting factor can be estimated for each age group. An expensive serological survey was conducted in Italy at the end of the first wave to determine the accurate overall count of new cases, but it could not provide a daily estimate. The present study introduces a cost-effective method to correct underreporting, which not only complements serological tests but yields daily time series. Our findings rely on the assumption that the cause-effect links between new cases and hospital admissions of the first and second waves were comparable. This assumption is quite plausible, given that no variants emerged and no vaccines were introduced during the early months of the pandemic.

\begin{table}[!ht]
\centering
\caption{{\bf Gains (percentage) and average time before hospitalization (days) with $2.5$th and $97.5$th percentiles, by age group.}}
\begin{tabular}{|c+c|c|c|c}
\hline 
 &  \textbf{Gain (\%)} 
 & \textbf{Delay (days)}
 & \textbf{Time constant (days)}
 \\ \thickhline
 $00 - 09$ & 1.76 (1.73, 1.79) & -10 (-10, -8) & 4.3 (2.5, 5.2) \\ \hline
$10 - 19$ & 0.68 (0.67, 0.69) & -10 (-10, -9) & 5.4 (3.9, 6.3) \\ \hline
$20 - 29$ & 1.44 (1.41, 1.45) & -10 (-10, -8) & 5.3 (3.1, 6.2) \\ \hline
$30 - 39$ & 2.3 (2.26, 2.34) & -10 (-10, -9) & 5.2 (3.8, 6.1) \\ \hline
$40 - 49$ & 3.19 (3.17, 3.2) & -10 (-10, -10) & 5.5 (5.3, 5.8) \\ \hline
$50 - 59$ & 5.72 (5.68, 5.75) & -7 (-8, -6.5) & 5.3 (4.7, 6.3) \\ \hline
$60 - 69$ & 11.98 (11.87, 12.05) & -7 (-8, -7) & 5.7 (5.4, 6.9) \\ \hline
$70 - 79$ & 22.54 (22.39, 22.66) & -6 (-6.5, -6) & 5.2 (4.9, 5.8) \\ \hline
$80 - 89$ & 30.62 (30.3, 30.83) & -8 (-9, -8) & 8.2 (8, 9.2) \\ \hline
$90 - 99$ & 25.06 (24.66, 25.33) & -10 (-10, -9) & 12.8 (11.1, 13.5) \\ \hline

\end{tabular}
\begin{flushleft} The gains, expressed as a percentage, and the average times before hospitalization with their $2.5$th and $97.5$th percentiles, vary greatly depending on the age group and reflect the differences in illness severity across different age groups. Both parameters appear to be estimated with good precision.
\end{flushleft}
\label{table1}
\end{table}

\begin{table}[!ht]
\centering
\caption{
{\bf Integrals of new positives (thousands).}}
\begin{tabular}{|c+c|c|c|c|}
\hline
 & \textbf{ISS} 
 & \textbf{Reconstructed} 
 & \textbf{U.C.}  
 \\ \thickhline


$00 - 09$ & 2 & 23 (22, 24) & 11.5 \\ \hline
$10 - 19$ & 4 & 47 (45, 48) & 11.75 \\ \hline
$20 - 29$ & 13 & 102 (100, 105) & 7.846 \\ \hline
$30 - 39$ & 18 & 139 (136, 142) & 7.722 \\ \hline
$40 - 49$ & 31 & 221 (219, 223) & 7.129 \\ \hline
$50 - 59$ & 42 & 236 (233, 238) & 5.619 \\ \hline
$60 - 69$ & 31 & 132 (131, 133) & 4.258 \\ \hline
$70 - 79$ & 33 & 89 (88, 90) & 2.697 \\ \hline
$80 - 89$ & 41 & 63 (63, 64) & 1.537 \\ \hline
$90 - 99$ & 19 & 21 (21, 22) & 1.105 \\ \hline
All ages & 234 & 1073 (1058 - 1089) & 4.585 \\ \hline
\end{tabular}
\begin{flushleft} The first two columns of the table respectively display the sum of the new cases recorded by ISS and reconstructed by the regularized deconvolution model (with their $2.5$-th and $97.5$-th percentiles). The last column shows the underestimation factor.
\end{flushleft}
\label{table2}
\end{table}

\section*{Conclusion}
The present paper highlights the critical importance of accurate data in controlling a pandemic outbreak.  The COVID-19 pandemic caught the healthcare and bureaucratic systems off-guard, resulting in severe underreporting of infections during the first wave in Italy. The lack of accurate data hampers the retrospective assessment of nonpharmacological interventions and the estimation and validation of epidemiological models.  The approach proposed in this paper, based on system identification and regularization for inverse problem, is valuable under two respects. First, it offers a valuable tool for analyzing the human and health costs of COVID-19 over a given period. Moreover, it enables a quantitative correction of underreporting. Our study shows that cause-and-effect relationships between numbers of infections and hospitalization admissions can be predicted with a high degree of reliability. This feature may prove critical to better respond to future epidemics in terms of safeguarding economic and health systems and implementing vaccine plans. In fact, the nature of the methodology is not strictly linked to the nature of COVID-19 or to one particular of its strains, although Alpha and Omicron variants are expected to require an update of the kernel function. One option is to conduct a piecewise-constant analysis, discontinuously studying each period associated with a particular variant. An alternative solution and a potential future research direction would be to replace the exponential parametric model with a time-varying impulse response. This approach can be modeled using identification techniques such as kernel methods \cite{gdn}, which would allow the model parameters to be time-dependent functions and not constrained to be constant over short periods.


\section*{Supporting information}



{\centering
\includegraphics[width=1\textwidth]{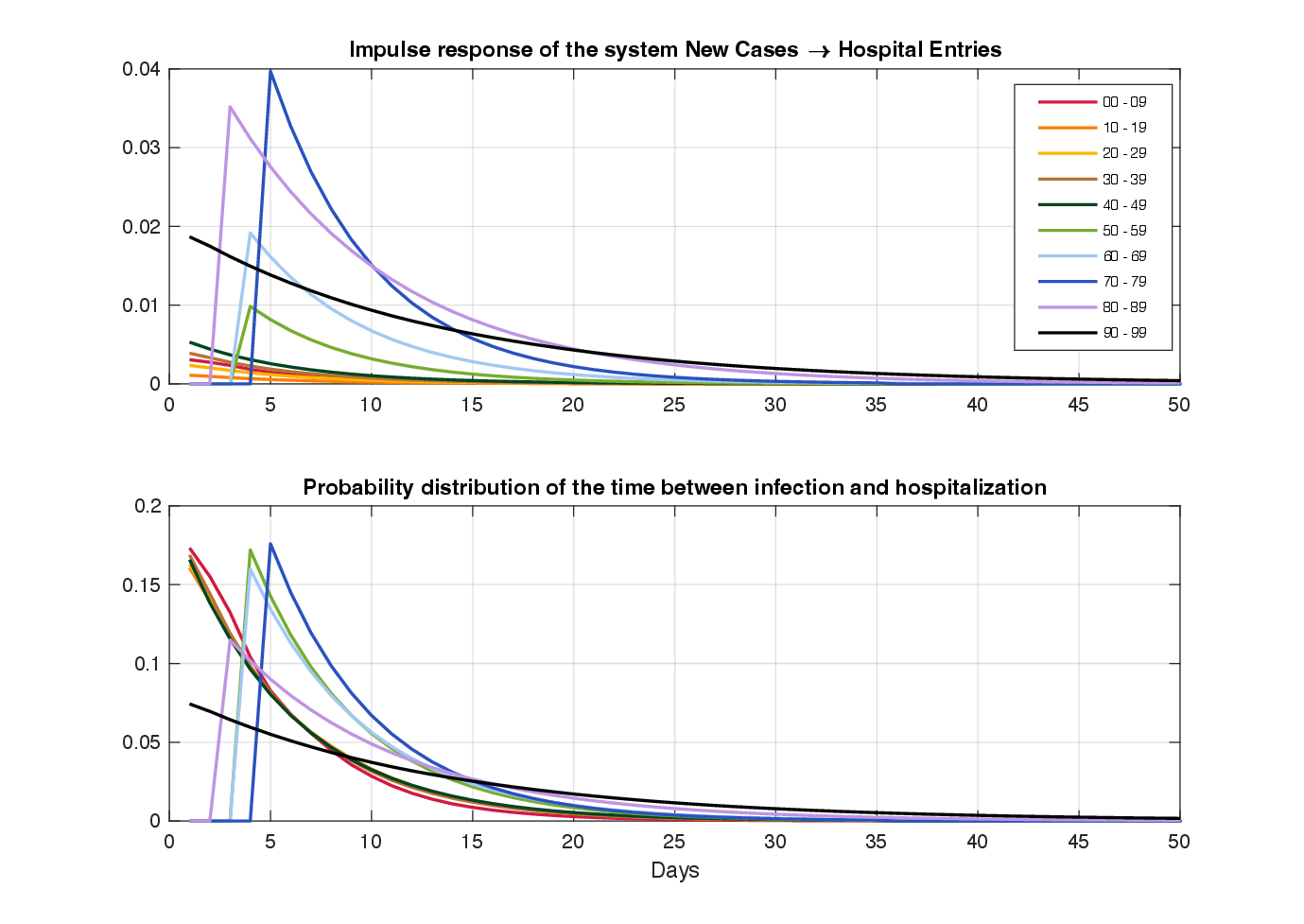}}
\label{S1Fig}
{\bf Fig S1. Comparison between impulse responses from different age groups.} The upper panel shows the ten different impulse responses. The lower panel of the figure shows ten different probability distributions of elapsed time from positivity. Among these distributions, the age groups up to 49 years old share a similar pattern, with a shorter mean compared to the three age groups between 50 and 79 years old. The last two age groups exhibit the slowest decay rates.



%
{\centering \includegraphics[width=1\textwidth]{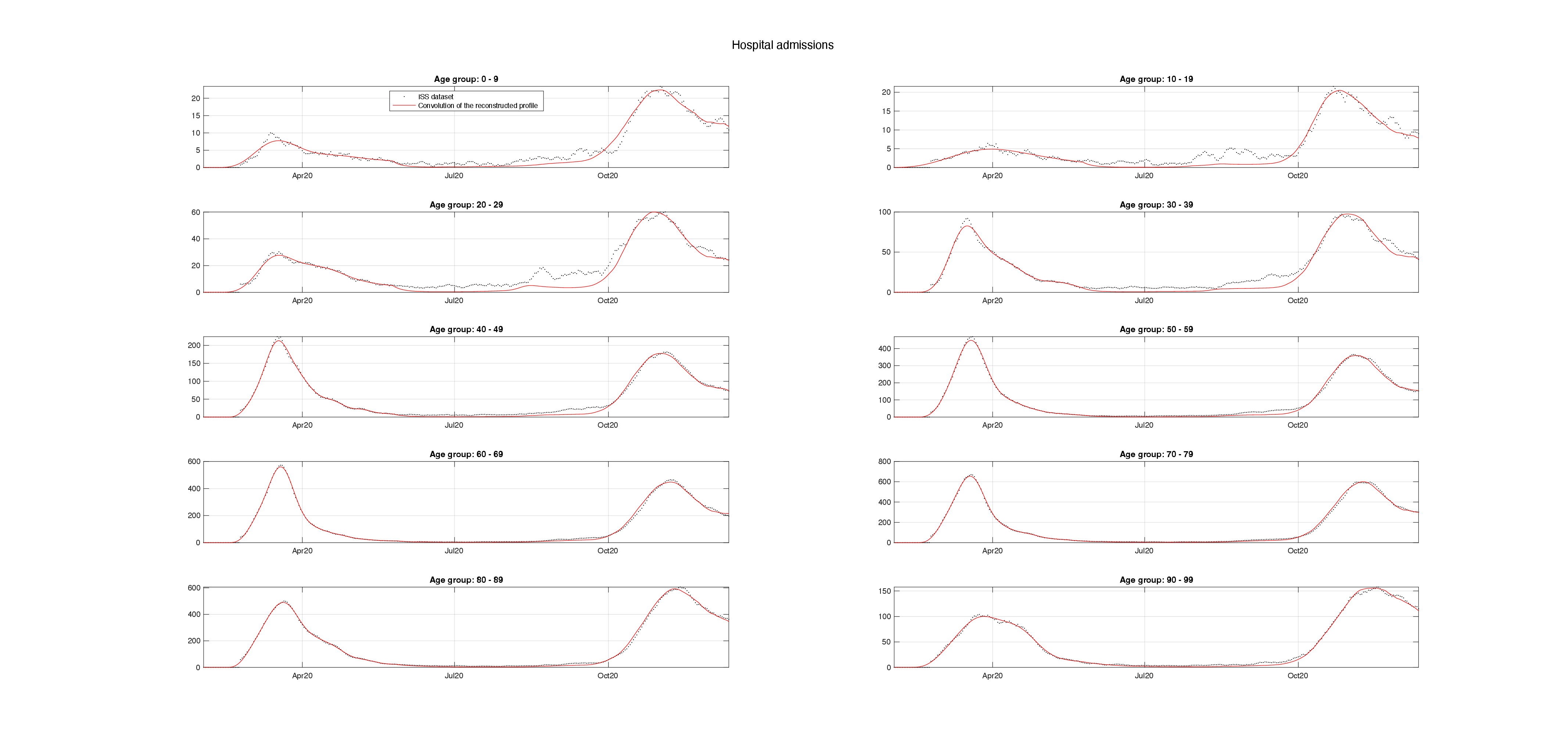}}
\label{S2Fig}
{\bf 
Fig S2. Reconvoluted dataset for each age group.} The hospital entry predictions for all age groups are presented. The model fits well, except for the younger age groups that, however, account for a smaller proportion of hospitalizations. Conversely, the fit for older age groups is satisfactory during the whole period.
%






\section*{Funding information}
This research was supported by EU funding within the NextGenerationEU-MUR PNRR Extended Partnership initiative on Emerging Infectious Diseases (Project no. PE00000007, INF-ACT).

\section*{Author contributions}
Simone Milanesi and Giuseppe De Nicolao have equally contributed in Conceptualisation, Methodology, Validation, and Writing. Data curation and visualization were  performed by Simone Milanesi and revised by Giuseppe De Nicolao.

\section*{Code and data availability statement}
 Code is available from the corresponding author upon request. Italian new cases and hospital admissions data are downloadable from \cite{sito_dati}.


%
%
%

\end{document}